\PassOptionsToPackage{unicode}{hyperref}
\PassOptionsToPackage{hyphens}{url}
\PassOptionsToPackage{dvipsnames,svgnames,x11names}{xcolor}
\documentclass[
]{article}

\usepackage{amsmath,amssymb}
\usepackage{iftex}
\ifPDFTeX
  \usepackage[T1]{fontenc}
  \usepackage[utf8]{inputenc}
  \usepackage{textcomp} % provide euro and other symbols
\else % if luatex or xetex
  \usepackage{unicode-math}
  \defaultfontfeatures{Scale=MatchLowercase}
  \defaultfontfeatures[\rmfamily]{Ligatures=TeX,Scale=1}
\fi
\usepackage{lmodern}
\ifPDFTeX\else  
\fi
\IfFileExists{upquote.sty}{\usepackage{upquote}}{}
\IfFileExists{microtype.sty}{% use microtype if available
  \usepackage[]{microtype}
  \UseMicrotypeSet[protrusion]{basicmath} % disable protrusion for tt fonts
}{}
\makeatletter
\@ifundefined{KOMAClassName}{% if non-KOMA class
  \IfFileExists{parskip.sty}{%
    \usepackage{parskip}
  }{% else
    \setlength{\parindent}{0pt}
    \setlength{\parskip}{6pt plus 2pt minus 1pt}}
}{% if KOMA class
  \KOMAoptions{parskip=half}}
\makeatother
\usepackage{xcolor}
\makeatletter
\ifx\paragraph\undefined\else
  \let\oldparagraph\paragraph
  \renewcommand{\paragraph}{
    \@ifstar
      \xxxParagraphStar
      \xxxParagraphNoStar
  }
  \newcommand{\xxxParagraphStar}[1]{\oldparagraph*{#1}\mbox{}}
  \newcommand{\xxxParagraphNoStar}[1]{\oldparagraph{#1}\mbox{}}
\fi
\ifx\subparagraph\undefined\else
  \let\oldsubparagraph\subparagraph
  \renewcommand{\subparagraph}{
    \@ifstar
      \xxxSubParagraphStar
      \xxxSubParagraphNoStar
  }
  \newcommand{\xxxSubParagraphStar}[1]{\oldsubparagraph*{#1}\mbox{}}
  \newcommand{\xxxSubParagraphNoStar}[1]{\oldsubparagraph{#1}\mbox{}}
\fi
\makeatother

\usepackage{longtable,booktabs,array}
\usepackage{calc} % for calculating minipage widths
\usepackage{etoolbox}
\makeatletter
\patchcmd\longtable{\par}{\if@noskipsec\mbox{}\fi\par}{}{}
\makeatother
\IfFileExists{footnotehyper.sty}{\usepackage{footnotehyper}}{\usepackage{footnote}}
\makesavenoteenv{longtable}
\usepackage{graphicx}
\makeatletter
\def\maxwidth{\ifdim\Gin@nat@width>\linewidth\linewidth\else\Gin@nat@width\fi}
\def\maxheight{\ifdim\Gin@nat@height>\textheight\textheight\else\Gin@nat@height\fi}
\makeatother
\setkeys{Gin}{width=\maxwidth,height=\maxheight,keepaspectratio}
\makeatletter
\def\fps@figure{htbp}
\makeatother
\NewDocumentCommand\citeproctext{}{}

\makeatletter
 \let\@cite@ofmt\@firstofone
 \def\@biblabel#1{}
 \def\@cite#1#2{{#1\if@tempswa , #2\fi}}
\makeatother
\newlength{\cslhangindent}
\newlength{\csllabelwidth}
\newenvironment{CSLReferences}[2] % #1 hanging-indent, #2 entry-spacing
 {\begin{list}{}{%
  \setlength{\itemindent}{0pt}
  \setlength{\leftmargin}{0pt}
  \setlength{\parsep}{0pt}
  \ifodd #1
   \setlength{\leftmargin}{\cslhangindent}
   \setlength{\itemindent}{-1\cslhangindent}
  \fi
  \setlength{\itemsep}{#2\baselineskip}}}
 {\end{list}}
\usepackage{calc}

\newcommand{\CSLLeftMargin}[1]{\parbox[t]{\csllabelwidth}{\strut#1\strut}}
\newcommand{\CSLRightInline}[1]{\parbox[t]{\linewidth - \csllabelwidth}{\strut#1\strut}}

\usepackage[noblocks]{authblk}

\usepackage{physics}
\usepackage{mathcal}
\usepackage{mathtools}
\usepackage{braket}
\makeatletter
\@ifpackageloaded{caption}{}{\usepackage{caption}}
\AtBeginDocument{%
\ifdefined\contentsname
  \renewcommand*\contentsname{Table of contents}
\else
  \newcommand\contentsname{Table of contents}
\fi
\ifdefined\listfigurename
  \renewcommand*\listfigurename{List of Figures}
\else
  \newcommand\listfigurename{List of Figures}
\fi
\ifdefined\listtablename
  \renewcommand*\listtablename{List of Tables}
\else
  \newcommand\listtablename{List of Tables}
\fi
\ifdefined\figurename
  \renewcommand*\figurename{Figure}
\else
  \newcommand\figurename{Figure}
\fi
\ifdefined\tablename
  \renewcommand*\tablename{Table}
\else
  \newcommand\tablename{Table}
\fi
}
\@ifpackageloaded{float}{}{\usepackage{float}}
\floatstyle{ruled}
\@ifundefined{c@chapter}{\newfloat{codelisting}{h}{lop}}{\newfloat{codelisting}{h}{lop}[chapter]}
\floatname{codelisting}{Listing}

\makeatother
\makeatletter
\@ifpackageloaded{caption}{}{\usepackage{caption}}
\@ifpackageloaded{subcaption}{}{\usepackage{subcaption}}
\makeatother

\ifLuaTeX
  \usepackage{selnolig}  % disable illegal ligatures
\fi
\usepackage{bookmark}

\IfFileExists{xurl.sty}{\usepackage{xurl}}{} % add URL line breaks if available
\hypersetup{
  pdftitle={Horizon-Driven Expansion from Hawking-Like Radiation: A Curvature-Coupled Cosmological Model},
  pdfauthor={Aman Singh},
  colorlinks=true,
  linkcolor={blue},
  filecolor={Maroon},
  citecolor={Blue},
  urlcolor={Blue},
  pdfcreator={LaTeX via pandoc}}

\title{Horizon-Driven Expansion from Hawking-Like Radiation: A
Curvature-Coupled Cosmological Model}

\author{Aman Singh}

\affil[aff-1]{University of Oxford, Oxford Research Centre, Oxford, UK}

\date{}
\begin{document}
\maketitle

\subsection{Abstract}\label{abstract}

We propose a cosmological model in which the expansion of the universe
is driven by a Hawking-like influx of energy across the cosmological
horizon, rather than from a fixed cosmological constant. In place of a
cosmological constant, we introduce source terms in the Friedmann and
continuity equations that couple horizon curvature to matter and
radiation densities. At high curvature (large Hubble parameter \(H\)),
this influx strongly replenishes matter and radiation, slowing their
adiabatic dilution. As curvature diminishes, the influx weakens,
smoothly transitioning into standard radiation- or matter-dominated
eras. This mechanism naturally suppresses spatial curvature without
requiring an inflationary phase. It may also produce
near-scale-invariant fluctuations via slowly varying horizon
thermodynamics.

\subsection{1. Introduction}\label{introduction}

The standard cosmological paradigm describes the evolution of the
universe through an initial period of inflation, followed by radiation
and matter domination, and culminating in a late-time acceleration
attributed to a cosmological constant \(\Lambda\)~{[}1,2{]}. While the
\(\Lambda\)CDM framework provides an excellent fit to a wide range of
observations, it leaves fundamental theoretical issues unresolved. Chief
among these are the origin and unnaturally small observed value of
\(\Lambda\), the fine-tuning of inflaton potentials necessary for
inflation, and the special initial conditions apparently required to
trigger inflation. Furthermore, persistent tensions between locally
measured values of the Hubble constant \(H_0\) and those inferred from
the cosmic microwave background (CMB) suggest that the standard model
may be incomplete~{[}3,4{]}.

Recent improvements in observational precision, particularly from baryon
acoustic oscillations (BAO) surveys and the Dark Energy Spectroscopic
Instrument (DESI) project~{[}5,6{]}, increasingly point to the
possibility that the dark energy component may be dynamical rather than
strictly constant. An unchanging cosmological constant, while simple,
may no longer fully describe the observed acceleration history. Although
many dynamical dark energy models have been proposed---including
k-essence~{[}7{]} and running vacuum scenarios~{[}8{]}---the
observational data until recently did not strongly necessitate such
complexity. With new data now favoring greater flexibility in the
expansion history, it is timely to revisit and extend the foundational
assumptions of cosmology.

A compelling alternative direction emerges from horizon thermodynamics,
extending analogies with black hole physics~{[}9{]}. If the cosmological
horizon possesses a Hawking-like temperature \(T_H \sim H\), it may emit
quanta into the observable universe. At early times, when the Hubble
rate \(H\) is large, the associated particle influx could contribute
significantly to the energy budget, dynamically modifying the universe's
expansion without requiring an inflationary phase confined to the
observable patch. As \(H\) decreases with expansion, the emission rate
would naturally diminish, asymptotically recovering standard matter- and
radiation-dominated behavior.

In this work, we propose a cosmological model in which the
horizon-sourced influx directly feeds the matter and radiation energy
densities, with each created quantum joining the appropriate component
and subsequently redshifting according to conventional laws. This
differs from adding a new independent fluid; instead, it modifies the
evolution of existing species in a dynamically consistent manner. The
result is an evolving expansion history where the universe's early
curvature can be damped by the accumulation of horizon-induced matter
and radiation, leading naturally toward flatness without the need for a
dedicated inflationary phase.

In the following sections, we develop the theoretical structure of this
horizon-driven influx model, describe its implementation via modified
continuity equations, and explore its observational implications for the
cosmic expansion history. We discuss how this framework may also
naturally generate near-scale-invariant perturbations and provide a
pathway toward resolving persistent cosmological tensions.

\subsection{2. Horizon-Driven Influx and Modified Continuity
Equations}\label{horizon-driven-influx-and-modified-continuity-equations}

In the standard cosmological model, the expansion of the universe is
governed by the Friedmann equations, with energy densities for matter,
radiation, and a cosmological constant evolving according to
well-established continuity relations. In the model proposed here, we
introduce an additional physical effect: a curvature-driven influx of
energy sourced by the cosmological horizon itself. This influx alters
the standard evolution of matter and radiation by acting as a continuous
source, dynamically modifying the energy budget as the universe expands.

The guiding principle is that the cosmological horizon, possessing an
effective Hawking temperature \(T_H \sim H\), emits quanta into the
observable universe. These quanta contribute to both the matter and
radiation energy densities, depending on the horizon temperature at a
given epoch. At very early times, when \(H\) and thus \(T_H\) are large,
the emission may include massive particles as well as radiation. As the
universe expands and \(H\) decreases, the temperature falls, favoring
radiation (photon and neutrino) production. The net effect is a
continuous injection of energy that gradually diminishes over time as
the curvature (and thus the emission rate) declines.

Rather than introducing a new independent fluid, we model this influx as
modifying the continuity equations of matter and radiation themselves.
Specifically, we modify the standard conservation laws by introducing
source terms \(Q_m(H)\) and \(Q_r(H)\):

\[
\frac{d\rho_m}{dt} + 3H \rho_m = Q_m(H),
\]

\[
\frac{d\rho_r}{dt} + 4H \rho_r = Q_r(H),
\]

where \(\rho_m\) and \(\rho_r\) are the matter and radiation energy
densities, respectively. The source terms represent the rates at which
horizon-sourced energy is deposited into the matter and radiation
components.

The total expansion dynamics remain governed by the Friedmann equation:

\[
H^2 = \frac{8\pi G}{3} \left( \rho_m + \rho_r \right).
\]

The total source term \(Q_{\text{tot}}(H)\), defined as
\(Q_{\text{tot}}(H) = Q_m(H) + Q_r(H)\), encapsulates the total energy
influx from the horizon. Motivated by dimensional considerations and the
analogy to horizon radiation, we assume a simple scaling form:

\[
Q_{\text{tot}}(H) = \alpha H^3,
\]

where \(\alpha\) is a small positive constant parameterizing the
efficiency of energy emission relative to the expansion rate. This form
arises naturally by noting that a horizon with temperature
\(T_H \sim H\) has an energy density scaling as \(T_H^4 \sim H^4\), and
an energy flux per unit area scaling as \(T_H^3 \sim H^3\).

To partition the influx between matter and radiation, we introduce a
dimensionless function \(f_m(z)\), depending on redshift \(z\),
describing the fraction of the horizon-sourced energy that converts into
matter:

\[
Q_m(H, z) = f_m(z) \, Q_{\text{tot}}(H),
\]

\[
Q_r(H, z) = \left( 1 - f_m(z) \right) Q_{\text{tot}}(H).
\]

At early times, when the horizon temperature is high, \(f_m(z)\) may be
close to unity, favoring matter production. As the universe cools,
\(f_m(z)\) should gradually decrease, favoring radiation production. For
simplicity, we adopt a phenomenological form such as:

\[
f_m(z) = \frac{1}{1 + (1+z)/z_{\text{eq}}},
\]

where \(z_{\text{eq}}\) is a parameter setting the redshift at which
matter and radiation production are comparable.

Together, these modified continuity equations define a self-contained
system for the evolution of \(\rho_m(z)\) and \(\rho_r(z)\), and hence
for the Hubble parameter \(H(z)\). By numerically integrating this
system, we can study the impact of horizon-sourced energy on the
expansion history, curvature evolution, and structure formation.

We next derive the key properties of this model, including the
asymptotic behavior at early and late times, and outline its observable
consequences.

We solve the system by writing
\(\dot{\rho} = \frac{d\rho}{dz}\frac{dz}{dt}\) and
\(\frac{dz}{dt}=-(1+z)H\). Thus

\[
\frac{d\rho_m}{dz} + \frac{3\,\rho_m}{1+z} 
= \frac{Q_m(H)}{H\,(1+z)},
\]

\[
\frac{d\rho_r}{dz} + \frac{4\,\rho_r}{1+z} 
= \frac{Q_r(H)}{H\,(1+z)}
\]

leading to

\[
\frac{d\rho_m}{dz} = \dot{\rho}_m \cdot \frac{dz}{dt} = \left[-3H\rho_m + Q_m\right] \cdot \left[-\frac{1}{H(1+z)}\right] = \frac{1}{H(1+z)} \left(3H\rho_m - Q_m\right)
\]

\[
H^2(z)=\frac{8\pi G}{3}\,\bigl[\rho_m(z)+\rho_r(z)\bigr].
\]

This forms a \textbf{coupled set} in \(\rho_m(z)\) and \(\rho_r(z)\),
with \(H(z)\) found algebraically.

\begin{center}\rule{0.5\linewidth}{0.5pt}\end{center}

\subsection{3. Flatness Without
Inflation}\label{flatness-without-inflation}

At large \(z\), \(H\) is large, so \(Q_{\mathrm{tot}}(H)\) strongly
raises \(\rho_m\) or \(\rho_r\). This keeps \(H\) from dropping too
fast, boosting \(aH\) and thereby suppressing \(\Omega_k=-\,k/(aH)^2\).
Thus curvature is driven towards zero without an exponential
inflationary expansion. Once curvature is sufficiently diluted, \(H\) is
smaller, so the horizon influx \(Q_{\mathrm{tot}}(H)\) diminishes.

In most models of inflation, one must arrange ``slow-roll'' and then
exit from the inflaton potential. Here, the horizon injection
automatically subsides as curvature falls (reducing horizon
temperature). Hence, the model seamlessly reverts to the usual
matter/radiation continuity once \(Q_m(H)\) and \(Q_r(H)\) become
negligible. The universe emerges with asymptotic flatness.

\begin{center}\rule{0.5\linewidth}{0.5pt}\end{center}

\subsection{4. Horizon-Sourced
Fluctuations}\label{horizon-sourced-fluctuations}

A hallmark of inflation is near-scale-invariant fluctuations from vacuum
modes freezing on super-horizon scales. In our approach, fluctuations
are seeded at the horizon scale by continuous horizon emission at
temperature \(T_H \approx H/(2\pi)\). Provided \(H\) changes slowly over
relevant e-folds, modes of different scales see nearly the same
amplitude, producing near-scale-invariance.

The observed amplitude of primordial curvature perturbations, inferred
from cosmic microwave background measurements, is approximately
\(\Delta_{\mathcal{R}}^2 \sim 2.1 \times 10^{-9}\)~{[}4{]}. In standard
inflationary theory, this amplitude arises from quantum fluctuations of
the inflaton field. In the present model, by contrast, perturbations
originate from thermal fluctuations of Hawking-like radiation sourced by
the horizon.

The characteristic scale of horizon fluctuations is set by the effective
horizon temperature \(T_H \sim H / (2\pi)\), where \(H\) is the Hubble
parameter during the early epoch of strong influx. The corresponding
dimensionless power spectrum is expected to scale as

\[
\Delta_{\mathcal{R}}^2 \sim \left( \frac{H}{2\pi M_{\mathrm{Pl}}} \right)^2,
\]

where \(M_{\mathrm{Pl}} = (8\pi G)^{-1/2}\) is the reduced Planck mass.
This relation assumes that fluctuations are predominantly thermal and
that no slow-roll suppression occurs, consistent with the absence of an
inflaton field in the horizon-driven model.

Solving for the required Hubble parameter yields

\[
H \approx 2\pi M_{\mathrm{Pl}} \sqrt{\Delta_{\mathcal{R}}^2}.
\]

Substituting numerical values, one finds

\[
H \sim 7 \times 10^{14}\,\mathrm{GeV}.
\]

Thus, if the Hubble parameter during the primordial
perturbation-generating epoch was of order
\(10^{14}\)--\(10^{15}\,\mathrm{GeV}\), the thermal fluctuations induced
by horizon-sourced radiation would naturally produce curvature
perturbations with the observed amplitude. This scale is compatible with
the expectation that the horizon curvature was extremely large at early
times, consistent with the general picture of a curvature-driven influx
model.

Importantly, this mechanism allows for a nearly scale-invariant spectrum
provided the horizon temperature evolves slowly relative to the Hubble
expansion. The precise tilt of the spectrum, and potential deviations
from exact scale invariance, depend on the rate of evolution of \(H(z)\)
during the influx-dominated era, and could be constrained by future
measurements of the spectral index \(n_s\).

\subsection{5. Observational
Consequences}\label{observational-consequences}

One can integrate \(\rho_m(z)\) and \(\rho_r(z)\) from some large
\(z_\mathrm{start}\) to \(z=0\). The fraction function \(f_m(H)\) sets
how matter vs.~radiation creation transitions as \(T_H\) drops below
certain mass thresholds. Here in figure 1, \(H(z)\) is tested against
BAO, supernova data, and CMB distance measures

\begin{figure}[H]

{\centering \includegraphics{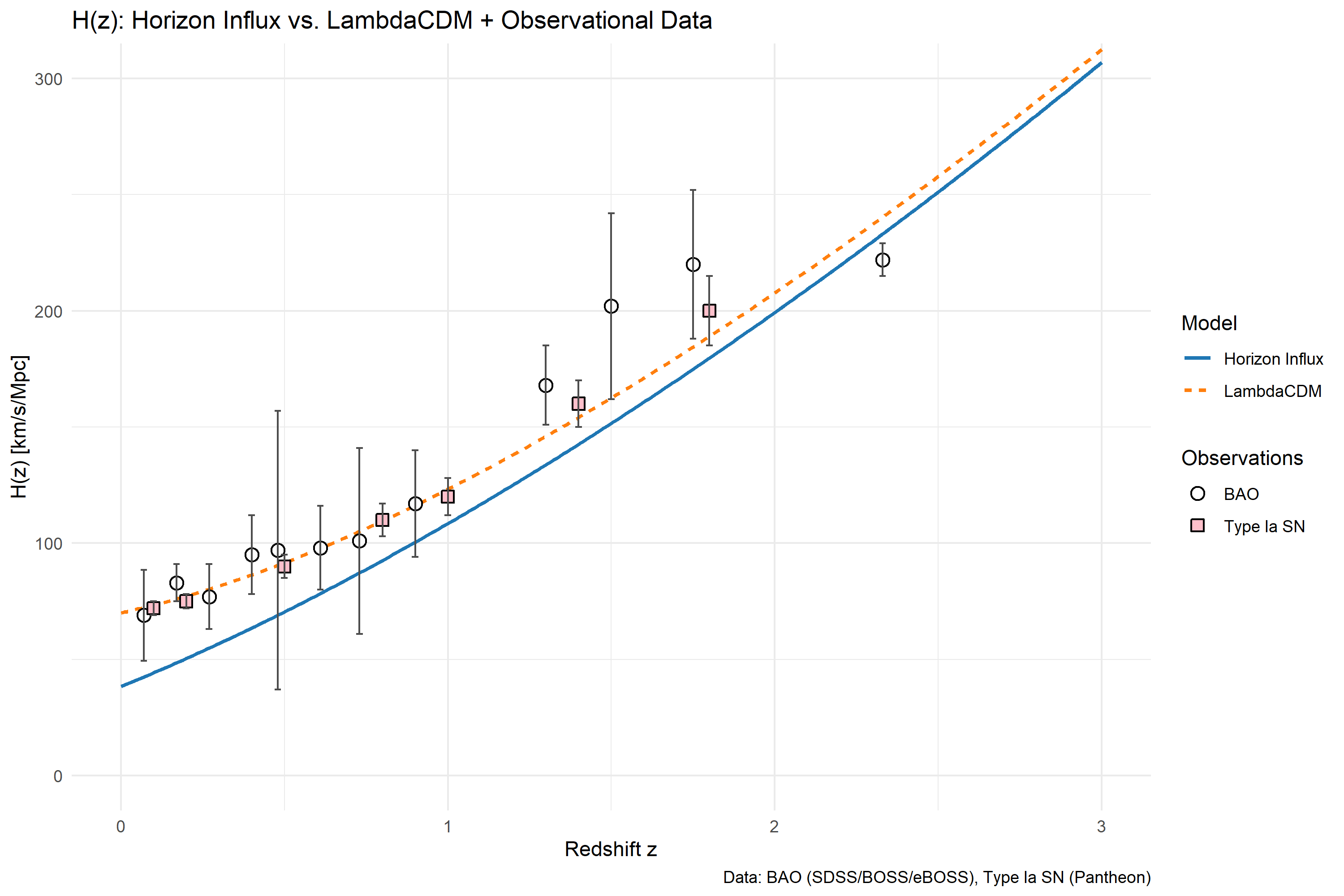}

}

\caption{Standard Lambda CDM model shown by the (dashed orange curve)
compared with the predicted horizon influx model (blue curve).
Superimposed observational data from BAO, type 1a supernovae shown.}

\end{figure}%

The Horizon Influx model remains consistent with observational
constraints but provides a greater number of degrees of freedom enabling
a level of fine tuning. This suggests that the Horizon Influx framework
offers a viable alternative to LambdaCDM, particularly given its
physically motivated curvature-dependent mechanism for cosmic
acceleration, potentially providing an exploratory path to resolving
tensions in cosmological measurements.

\begin{center}\rule{0.5\linewidth}{0.5pt}\end{center}

\subsection{6. Conclusion}\label{conclusion}

We introduced and analyzed a cosmological framework wherein the
expansion of the universe is driven by a curvature-dependent energy
influx originating from the cosmological horizon, replacing the
conventional cosmological constant. The proposed model modifies the
Friedmann and continuity equations through curvature-dependent source
terms, providing a continuous energy input into matter and radiation
components. At early times, characterized by high curvature, this influx
substantially mitigates the dilution of matter and radiation densities,
thereby naturally reducing spatial curvature without requiring an
inflationary epoch.

The curvature-dependent influx in this approach decreases progressively
as curvature reduces during cosmic expansion, smoothly transitioning
into conventional matter- and radiation-dominated epochs. The
self-regulating nature of this mechanism ensures the emergence of a
nearly flat universe through standard expansion dynamics rather than a
period of inflationary expansion. Furthermore, the thermodynamic
properties associated with horizon emission offer a mechanism capable of
generating near-scale-invariant perturbations, provided the horizon
temperature evolves gradually during key epochs.

The horizon-driven influx model offers a physically motivated
alternative explanation for cosmic acceleration and structure formation
without relying extensively on additional scalar fields or arbitrary
parameters. By incorporating principles analogous to horizon
thermodynamics, this model provides an alternative approach to
cosmological dynamics.

\begin{center}\rule{0.5\linewidth}{0.5pt}\end{center}

\section*{References}\label{bibliography}
\addcontentsline{toc}{section}{References}

\phantomsection\label{refs}

\end{document}